\documentclass[a4paper,11pt]{article}

\usepackage{authblk}
\usepackage[a4paper]{geometry}
\usepackage{color}
\usepackage{graphicx}
\usepackage{amsmath}
\usepackage{amsfonts}
\usepackage{amsthm}
\usepackage{amscd}
\usepackage{epsfig}
\usepackage{amssymb}
\usepackage{slashed}
\usepackage{tabularx}
\usepackage{calligra}
\usepackage{enumerate}
\usepackage{hyperref}
\usepackage{float}
\usepackage{stackrel}
\usepackage[T1]{fontenc}
\usepackage{longtable}
\usepackage{amsthm}
\usepackage{mathrsfs}
\usepackage{verbatim} 
\usepackage{fancyhdr}
\usepackage{tikz}
\usepackage{tikz-cd}
\usetikzlibrary{matrix}
\usetikzlibrary{positioning}
\usetikzlibrary{decorations.pathmorphing,calc}
\usepackage[all]{xy}
\usepackage{units}
\usepackage{textcomp}
\usepackage{turnstile}
\usepackage{vmargin}
\usepackage{anysize}
\usepackage{ stmaryrd }
\usepackage{tikz-3dplot}
\usepackage{soul}

\usepackage{caption}
\usepackage{mnotes}

\setmargins{2,5cm}       
{1,5cm}                        
{16cm}                      
{23,42cm}                    
{10pt}                           
{1cm}                           
{0pt}                             
{2cm}                           

\theoremstyle{plain}

\numberwithin{equation}{section}

\def\d{{\rm d}}
\def\i{{\rm i}}

\def\ben{\begin{equation}}
\def\bea{\begin{eqnarray}}
\def\een{\end{equation}}
\def\eea{\end{eqnarray}}

\newcommand{\Log}{\text {Log}}

\begin{document}

\title{{\bf {$\alpha-$surfaces} {and global coordinates in} black hole spacetimes}}
\author[1]{Bernardo Araneda\footnote{Email: \texttt{bernardo.araneda@aei.mpg.de}}}
\author[2]{Bernard F. Whiting\footnote{Email: \texttt{bernard@phys.ufl.edu}}}
\affil[1]{Max-Planck-Institut f\"ur Gravitationsphysik (Albert-Einstein-Institut), \protect\\Am M\"uhlenberg 1, D-14476 Potsdam, Germany}
\affil[2]{Department of Physics, University of Florida, Gainesville, Florida, USA}

\date{\today}

\maketitle

\begin{abstract}
We present a discussion of the periodicity in imaginary time of maximally extended black hole spacetimes without reference to Euclidean manifolds.  As motivation, we first demonstrate our approach for the Rindler geometry in flat 
space before then applying it for the case of Schwarzschild time in the maximally extended Kruskal geometry.  One notable advantage of our approach is that it can be utilized in the Kerr case, again without reference to a Euclidean manifold, and without complexifying the angular momentum parameter, $a$.  One unusual feature of our application in the Kerr case is that{, even for the purely Lorentzian geometry, it is developed in terms of explicitly complex coordinates which are associated with distinct families of $\alpha-$surfaces}.  Moreover, coordinatization of these gives a single set of coordinates which cover the entire geometry outside the inner horizon.
\end{abstract}

\section{Introduction}
Periodicity in imaginary time for black hole spacetimes obviously implies considering some complexification of the spacetime coordinates, but does not require that one obtain a fully Euclidean section in the complexified manifold.  However, 
in the path integral approach to the quantization of gauge theories, the use of positive-definite (Euclidean) rather than Lorentzian manifolds can be motivated (see e.g. \cite{GHP1978}) by the observation that the path integral is better behaved for the former class of metrics. The extension of this method to the quantization of gravitational fields leads to the Euclidean approach to quantum gravity \cite{GHbook}, where a similar motivation can be given \cite{Hawking1979}. An additional feature in the gravity case is that, since the path integral includes geometries such as black holes, one faces difficulties in the evaluation of the gravitational action due to spacetime singularities \cite{GH1977}. The Euclidean approach circumvents this issue by evaluating the action on a positive-definite section of the complexified geometry that is complete and free of singularities (a ``gravitational instanton'' \cite{Hawking1976}).

The Euclidean section of the Schwarzschild solution can be obtained by simply replacing $t$ by ${-}\i\tau$ in Schwarzschild coordinates (where $\i^2=-1$), and declaring $\tau$ to be real. One then notices that the space is free of (conical) singularities if $\tau$ is interpreted as an angular variable with certain periodicity. When trying to do the same for the Kerr solution, one finds that the replacement $t\to{-}\i\tau$ (using now Boyer-Lindquist coordinates) produces a complex metric \cite{GH1977}, {\em unless} one simultaneously replaces the angular momentum parameter $a$ by $\i a'$ and declares $a'$ to be real; this then gives the Euclidean section of Kerr \cite{GH1979}. The singularity-free requirement leads to certain identifications for the $\tau$ and $\phi$ variables of the Euclidean section \cite{GH1979}.

The use of imaginary time can also be motivated from quantum-mechanical and thermodynamical considerations different to those given above (e.g., the Schr\"odinger and heat equations are related by a Wick rotation; more generally see \cite{Zee}). The approach to black hole thermodynamics from the Euclidean quantum gravity perspective would then seem to require finding real, positive-definite, complete and singularity-free sections of black hole spacetimes. However, it is perhaps hard to give a compelling physical argument for the replacement $a\to\i a'$ in the Kerr solution (see e.g. \cite{Witten, Brown}), since $a$ is a fixed parameter that is supposed to characterize the rotation of an astrophysical black hole. (In fact, the original derivation of the thermodynamical properties of rotating black holes in {\cite{GH1977}} uses directly the complex Kerr geometry.) 

\medskip
In this paper, we wish to show that periodicity in imaginary time, together with possible imaginary shifts in angular variables, can be derived from a unified treatment of the global, analytic, and ``Hermitian'' properties of both static and rotating black hole spacetimes, without any reference to Euclidean sections, replacements of the sort $a\to\i a'$, or explicit requirements of absence of conical singularities.  

The ``global'' and ``analytic'' properties are familiar in relativity; we are here referring to the fact that black holes have horizons that act as boundaries for geodesically incomplete regions, which leads {one }to look for analytic extensions across these boundaries and to the construction of a global black hole spacetime. As we highlight further below, the extension is constructed by exploiting the special structure of null geodesics of these black holes.

The ``Hermitian'' property is perhaps less known: it refers to the fact, discovered by Flaherty \cite{Flaherty0, Flaherty}, that black hole solutions in general relativity have the Lorentzian analog of an integrable almost-complex structure compatible with the metric.
This implies that any principal null geodesic in a black hole spacetime is a real ray in an {\em $\alpha-$surface} (or {\em twistor surface}): a 2-complex-dimensional, totally null, totally geodesic, surface lying in the complexified manifold \cite{Penrose1976, PR2}. 
As we will show, $\alpha-$surfaces provide special complex coordinate systems that: allow to extend the spacetime across future and past horizons, cover the entire geometry outside the inner horizon of Kerr, and lead to periodicity in imaginary time along with an imaginary shift in the azimuthal angle. 

\medskip
We began our investigation by using coordinates defined by geodesics.  However, for the Kerr case, they proved to be insufficient for the task.  Instead, it became apparent during the course of our work that the presence of families of $\alpha-$surfaces in the complexified geometries played a distinguished role in our approach.  Black hole horizons are totally geodesic null surfaces.  
Moreover, we will show that, in complexified black hole spacetimes, their horizons (future and past) lie within (distinct families of) $\alpha-$surfaces.  Although we focus on the Schwarzschild and Kerr solutions, our method can in principle be applied to any black hole spacetime that admits $\alpha-$surfaces. This includes \cite{AA} the complete Pleba\'nski-Demia\'nski family of solutions \cite{PD}, so one can add electromagnetics charges, cosmological constant, acceleration, and NUT parameters. These are all Petrov type D, but $\alpha-$surfaces turn out to also exist in black holes that are not algebraically special, such as Kerr-Sen and other solutions of supergravity \cite{AA}, and our procedure will also apply in these cases too.

In section {\ref{Rindler}, we illustrate} our method for the 2-dimensional ({2D}) Rindler spacetime{.  In \ref{Sec:Kruskal}}, 
we apply our construction to the Kruskal extension of Schwarzschild. In section \ref{sec:Kerr} we present our main results for Kerr. We conclude with a discussion in section \ref{sec:Discussion}. We include an appendix \ref{Appendix} with further details about complex structures in Lorentz signature.

\section{{Rindler}}
\label{Rindler}

\begin{figure}
\centering
\begin{tikzpicture}

\node (I)    at ( 4,0)   {I};
\node (I')   at (0,0)   {I$'$};
\node (II)  at (2, 1.5) {II};
\node (II')   at (2,-1.5) {II$'$};

\path 
   (I') +(0,2)  coordinate (Itop)
       +(0,-2) coordinate (Ibot)
       +(4,2) coordinate (Ileft)
       +(4,-2)  coordinate (Iright)
       +(2,0) coordinate (bs)
       ;

\draw (Ibot) -- 
		node[xshift=1.8cm, yshift=-0.2cm, sloped] {\tiny $\rho=0$, $t=\infty$} 
		node[xshift=1.8cm, yshift=0.2cm, sloped] {\scriptsize $\mathcal{H}^{\rm f}$} 
	(Ileft);
\draw (Itop) -- 
		node[xshift=1.8cm, yshift=0.2cm, sloped] {\tiny $\rho=0$, $t=-\infty$}  
		node[xshift=1.8cm, yshift=-0.2cm, sloped] {\scriptsize $\mathcal{H}^{\rm p}$}
		(Iright);
\draw[->] (-2,0) -- node[left]{$T$} (-2, 1);
\draw[->] (-2,0) -- node[below]{$X$} (-1, 0);
\draw (4.2,1.6) node[right] {\tiny $\rho={\rm const.}$} to[out=215, in=145] (4.2,-1.6);
\draw[latex-] (4,2) to[out=-30,in=180] (5,2.2) node[right]{$U=0$};
\draw[latex-] (4,-2) to[out=30,in=180] (5,-1.5) node[right]{$V=0$};

\end{tikzpicture}
\caption{Maximal extension of the Rindler wedge $\mathcal{R}={\rm I}\backslash(\mathcal{H}^{\rm f}\cup\mathcal{H}^{\rm p})$. A line $\rho={\rm const.}$ in ${\rm I}$ represents an observer undergoing uniform acceleration, since $\rho=UV=-T^2+X^2$.}
\label{Fig:Rindler}
\end{figure}
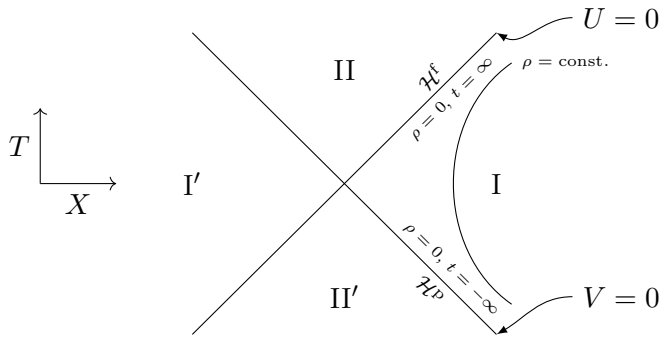

The basic intuition can be developed by looking at the {2D} Rindler spacetime, 
{with its metric in a form motivated by reference to the Schwarzschild metric.  
Consider, then, }{the Lorentzian metric
\begin{align}
 \d{s}'^2 = -f(r)\d{t}^2+\frac{\d{r}^2}{f(r)} \label{2dS}
\end{align}
where $f(r)=1-2M/r$, $M$ is a real positive constant, and $(t,r)\in\mathbb{R}\times(2M,\infty)$. The metric \eqref{2dS} has an apparent singularity at $r=2M$. Performing a Taylor expansion of $f(r)$ around $r=2M$, we have $f(r)\approx\frac{(r-2M)}{2M}$. Defining $\rho:=8M(r-2M)$ and $\alpha:=\frac{1}{4M}$, the metric \eqref{2dS} near $r=2M$ becomes
\begin{align}
 \d{s}^2 = -\alpha^2\,\rho\,\d{t}^2 + \frac{\d\rho^2}{4\rho}, \label{Rindler2}
\end{align}
where $(t,\rho)\in\mathbb{R}\times(0,\infty)$. The Lorentzian manifold with metric \eqref{Rindler2} and the above range of coordinates is the {2D} Rindler spacetime, $\mathcal{R}$. From the above we see that it is a  {2D} description of the near horizon geometry of the Schwarzschild solution.} The apparent singularity is now the region $\{\rho=0\}$. To analyze this region, we follow \cite[Sec. 6.4]{Wald} and introduce coordinates based on null geodesics. The null geodesics of \eqref{Rindler2} satisfy $-\alpha^2\rho\,\dot{t}^2+\frac{\dot{\rho}^2}{4\rho}=0$, so $\frac{\d{t}}{\d\rho} = \pm\frac{1}{2\alpha\rho}$, and the two signs give:
\begin{align}
\text{``outgoing'':} \quad t = +\tfrac{1}{2\alpha}\log|\rho| + u, 
\qquad\quad
\text{``ingoing'':} \quad t = -\tfrac{1}{2\alpha}\log|\rho| + v,
 \label{NGR}
\end{align}
where $u,v$ are constants, and for later convenience we have written $|\rho|$ even though $\rho>0$. Note that in the outgoing case, $t\to-\infty$ as $\rho\to0$, and in the ingoing case $t\to+\infty$ as $\rho\to0$. We use $(u,v)$ as coordinates:
\begin{align}
u = t-\tfrac{1}{2\alpha}\log|\rho|, \qquad v = t+\tfrac{1}{2\alpha}\log|\rho|. \label{zRindler}
\end{align}
$u={\rm const.}$ fixes an outgoing geodesic, and $v={\rm const.}$ fixes an ingoing geodesic. Defining
\begin{align}
 U = e^{-\alpha u}, \qquad V = e^{\alpha v}, \label{UVR}
\end{align}
the metric becomes
\begin{align}
\d{s}^2 = \d{U}\d{V}. \label{Rindler3}
\end{align}
Note that:
\begin{align}
\rho=UV \quad{\rm and}\quad t=\frac{1}{2\alpha}\log\left|\frac{V}{U}\right| \label{R-coords}
\end{align}
where we have dropped the $|\cdot|$ on $\rho$ to allow an anticipated smooth transition across $\rho=0$.  Setting:
\begin {align}
U=-T+X, \quad V=T+X, \quad{\rm we~get}\quad \d{s}^2 = -\d{T}^2+\d{X}^2, \label{R-ext}
\end{align}
so $\mathcal{R}$ is  locally isometric to {2D} Minkowski spacetime, $\mathbb{M}^2$. However, \eqref{UVR} implies that $U>0$, $V>0$, that is $-X<T<X$. Thus $\mathcal{R}$, which is known as the Rindler wedge, is only a portion of $\mathbb{M}^2$ and is geodesically incomplete. See fig. \ref{Fig:Rindler}.

From the behaviour of geodesics, we see that when $T\to \pm X$, we have $t\to \pm \infty$. $\mathcal{R}$ is then bounded by $\mathcal{H}\equiv\mathcal{H}^{\rm f}\cup\mathcal{H}^{\rm p}$, where $\mathcal{H}^{\rm f/p}=\{T=\pm X\}$ is the future/past ``horizon''.

From \eqref{Rindler3} one sees that the metric components are regular at $\mathcal{H}$, so the ``singularity'' $\{\rho=0\}$ is only an apparent singularity, and the spacetime can be extended across both horizons. This can be done by defining a larger spacetime with the same local form of the metric, eq. \eqref{Rindler3}, but where now the coordinate values $U<0$ and $V<0$ are allowed. We then define the regions
\begin{align*}
 {\rm I}=\{U\geq0, V\geq0\}, \quad  {\rm II}=\{U\leq0, V\geq0\},  \quad {\rm I'}=\{U\leq0, V\leq0\}, \quad  {\rm II'}=\{U\geq0, V\leq0\}.
\end{align*}
The resulting extended spacetime, ${\rm I}\sqcup{\rm II}\sqcup{\rm I'}\sqcup{\rm II'}$, is obviously $\mathbb{M}^2$.  Since we have dropped the $|\cdot|$ in the first of eqns (\ref{R-coords}), they can now be used without modification with $U,V$ negative.  Notice that $\rho$ is negative in regions ${\rm II,II'}$, and positive in ${\rm I,I'}$.

The timelike Killing vector $\frac{\partial}{\partial{T}}$ is global and provides a time-orientation for $\mathbb{M}^2$. With respect to this time-orientation, the Killing vector $\frac{\partial}{\partial{t}}$ associated to Rindler time is future-directed in ${\rm I}$ and past-directed in ${\rm I'}$, while it 
is spacelike in regions ${\rm II}$ and ${\rm II'}$.

How should we see the new regions ${\rm II, I', II'}$ from the perspective of the original Rindler coordinates \eqref{Rindler2}? From (\ref{R-coords}), each of the four regions ${\rm I, II, I', II'}$ has its own coordinates $(t,\rho)$, which can be used in (\ref{zRindler}). We can then set
\begin{align}
 U:= \begin{cases} e^{-\alpha u}, \quad {\rm I, II'}  \\ -e^{-\alpha u}, \quad {\rm I',II} \end{cases}, 
 \qquad
 V:= \begin{cases} e^{\alpha v}, \quad {\rm I,II}  \\ -e^{\alpha v}, \quad {\rm I',II'} \end{cases}
\end{align}
where $(u,v)$ are related to $(t,\rho)$ by \eqref{zRindler}.  
To understand better the extension beyond our original domain $\mathcal{R}$, we will introduce the principal value function ``$\Log(z)$'', analytic on the complex plane cut from $0$ to $-\infty$, undefined at the origin, with $\arg{z}\in(-\pi,\pi]$, rather than the more familiar ``$\log(\rho)$'', defined only for positive, real $\rho$.  Then we introduce new (generally complex) null coordinates:
\begin{align}
\tilde{u} = t-\tfrac{1}{2\alpha}\Log(\rho), \qquad \tilde{v} = t+\tfrac{1}{2\alpha}\Log(\rho). 
\label{zRind-tilde}
\end{align}
Note $L=\Log(\rho)$ receives an imaginary $\delta L=\i\pi$ jump when the sign of $\rho$ changes from positive to negative on moving from region ${\rm I}$ into regions ${\rm II}$ or ${\rm II'}$.  For $e^{\alpha \tilde{v}}$ to remain continuous across the future horizon $\mathcal{H}^{\rm f}$, and for $e^{-\alpha \tilde{u}}$ to remain continuous across the past horizon $\mathcal{H}^{\rm p}$ in passing out of region ${\rm I}$, it is then necessary that $t$ receive a $\delta t_{\rm I,II}=-\i\pi/(2\alpha)$ jump when passing into region ${\rm II}$, and a $\delta t_{\rm I,II'}=+\i\pi/(2\alpha)$ jump when passing into region ${\rm II'}$.  As a consequence, in region ${\rm II}$, $e^{-\alpha \tilde{u}}$ becomes $-e^{-\alpha u}={U|_{\rm II}}$, and in region ${\rm II'}$, $e^{\alpha \tilde{v}}$ becomes $-e^{\alpha v}={V|_{\rm II'}}$.  

Similarly, in passing from regions ${\rm II}$ or ${\rm II'}$ into region ${\rm I'}$, $L=\Log(\rho)$ will receive a relative $\delta L=-\i\pi$ jump, while $t$ will receive an additional $\delta t_{\rm II,I'}=-\i\pi/(2\alpha)$ jump when passing from region ${\rm II}$ into region ${\rm I'}$, so that $e^{\alpha \tilde{v}}$ becomes $-e^{\alpha v}={V|_{\rm I'}}$ while $e^{-\alpha \tilde{u}}$ remains $-e^{-\alpha u}={U|_{\rm I'}}$; alternatively, $t$ will receive an additional $\delta t_{\rm II',I'}=\i\pi/(2\alpha)$ jump when passing from region ${\rm II'}$ into region ${\rm I'}$, so that $e^{-\alpha \tilde{u}}$ becomes $-e^{-\alpha u}={U|_{\rm I'}}$ while $e^{\alpha \tilde{v}}$ remains $-e^{\alpha v}={V|_{\rm I'}}$.  

These last two results must be equivalent, so $t$ must have imaginary periodicity $\beta=2\pi/\alpha$ on the extended manifold, $\mathbb{M}^2$, i.e.:
\bea
t \sim t + \i\beta, \quad{\rm with}\quad \beta\equiv\frac{2\pi}{\alpha}.\label{R-per}
\eea
With the understanding we have thus developed for jumps in $t$ between neighbouring regions, ($\tilde{u},\tilde{v}$) become global, piecewise continuous coordinates, and the identifications:
\begin{align}
 \tilde{U} = e^{-\alpha \tilde{u}}, \qquad \tilde{V} = e^{\alpha \tilde{v}}, \label{UVR-t}
\end{align}
can be applied in all four regions simultaneously.

\section{{Schwarzschild, and its Kruskal Extension}}
\label{Sec:Kruskal}

This case is very similar to the Rindler example. Consider the Schwarzschild exterior $\mathcal{D}$:
\begin{align}
 \d{s}^2 = -f(r)\d{t}^2 +\frac{\d{r}^2}{f(r)} + r^2(\d\theta^2+\sin^2\theta\d\phi^2), 
 \qquad f(r):=1-\frac{2M}{r},
 \label{Schw}
\end{align}
where $(t,r,\theta,\phi)\in\mathbb{R}\times(2M,\infty)\times S^2$ and $M$ is a positive constant. To analyze the apparent singularity at $r=2M$, we again focus on null geodesics. 

We are not in {2D} anymore, so we have more freedom for the geodesics{, b}ut it turns out the geometry has ``preferred'' null geodesics.  Black hole spacetimes, including those of Schwarzschild and Kerr, are of Petrov type D, and have repeated principal null directions (PNDs), that we refer to as $\ell^a$ and $n^a$ below.  
A convenient null tetrad aligned to the PNDs is 
\begin{subequations}\label{PNDS}
\begin{align}
 \ell^{a}\partial_{a} &= \tfrac{1}{\sqrt{2f}}(\partial_{t}+f\partial_{r}), \label{ellS} \\ 
 n^{a}\partial_{a} &= \tfrac{1}{\sqrt{2f}}(\partial_{t}-f\partial_{r}), \label{nS} \\ 
 m^{a}\partial_{a} &= \tfrac{1}{r\sqrt{2}}(\partial_{\theta}+\tfrac{\i}{\sin\theta}\partial_{\phi}), \\
 \bar{m}^{a}\partial_{a} &= \tfrac{1}{r\sqrt{2}}(\partial_{\theta}-\tfrac{\i}{\sin\theta}\partial_{\phi}).
\end{align}
\end{subequations}
We see that only $\ell^a$ and $n^a$ have problems at the horizon, $r=2M$, so we can focus on the null geodesics associated to these vectors. 

{G}{iven a vector field $v=v^{a}\partial_{a}$ in $M$, an integral curve of $v$ is a curve $\mathbb{R}\to M$, $\tau\mapsto x^{a}(\tau)$ such that $v^{a}=\frac{\d{x}^{a}}{\d\tau}\equiv\dot{x}^{a}$. In Schwarzschild coordinates $(t,r,\theta,\phi)$, we then have $v = \dot{t} \, \partial_t + \dot{r} \, \partial_r + \dot\theta \, \partial_{\theta} + \dot\phi \, \partial_{\phi}$. The integral curves of $\ell^a$ and $n^a$ are the null geodesics of the PNDs. For $v^{a}=\ell^a$, from \eqref{ellS} we deduce that $\dot{t}=(2f)^{-1/2}$, $\dot{r}=(\frac{f}{2})^{1/2}$, $\dot\theta=0$ and $\dot\phi=0$. Similar expressions are deduced for $n^{a}$ from \eqref{nS}.} So we see that the principal null congruences satisfy
\begin{align}
 \frac{\d{t}}{\d{r}} = \pm\frac{1}{f}, \qquad \theta={\rm const.}, \qquad \phi={\rm const.}
\end{align}
where the $+$ sign corresponds to $\ell^a$ and gives outgoing geodesics, and $-$ corresponds to $n^{a}$ and gives ingoing geodesics. Generalizing \eqref{NGR}, the solutions are 
\begin{align}\label{tSchwPNDs} 
 \text{outgoing:} \,\, 
 \begin{cases}
 t = r + 2M\log\left| \tfrac{r}{2M} - 1 \right| + u \\
 \theta={\rm const.} \\ 
 \phi ={\rm const.}
 \end{cases},
 \qquad
 \text{ingoing:} \,\, 
 \begin{cases}
 t = - r - 2M\log\left| \tfrac{r}{2M} - 1 \right| + v \\
 \theta={\rm const.} \\
  \phi ={\rm const.}
 \end{cases}
\end{align}
where $u,v$ are constants, and for later purposes we have written $\left| \frac{r}{2M} - 1 \right|$ even though $r>2M$. Note that in the outgoing case, $t\to-\infty$ as $r\to2M$, and in the ingoing case $t\to+\infty$ as $r\to2M$, and all in/out-going geodesics are incomplete. As in Rindler \eqref{zRindler}, we use $u,v$ as coordinates:
\begin{align}\label{zSchw}
 u = t - r - 2M\log\left| \frac{r}{2M} - 1 \right|, \qquad 
 v = t+r + 2M\log\left| \frac{r}{2M} - 1 \right|. 
\end{align}
Defining
\begin{align}
 U=e^{-\kappa u}, \qquad V=e^{\kappa v}, \label{UVSchw}
\end{align}
where $\kappa:=\frac{1}{4M}$, we find that:
\bea
\left(\frac{r}{2M}-1\right){e}^{(r/2M)}=UV \quad{\rm and}\quad t=\frac{1}{2\kappa}\log\left|\frac{V}{U}\right|. \label{S-coords}
\eea
and the metric becomes:
\begin{align}
 \d{s}^2=\frac{32M^3}{r}e^{-\frac{r}{2M}}\d{U}\d{V} + r^2(\d\theta^2+\sin^2\theta\d\phi^2),
 \label{Kruskal}
\end{align}
where $r$ is viewed as a function $r(U,V)$ defined implicitly by \eqref{S-coords}.

\begin{figure}
\centering
\begin{tikzpicture}

\node (I)    at ( 4,0)   {I};
\node (I')   at (0,0)   {I$'$};
\node (II)  at (2, 1) {II};
\node (II')   at (2,-1) {II$'$};

\path 
   (I') +(0,2)  coordinate (Itop)
       +(0,-2) coordinate (Ibot)
       +(4,2) coordinate (Ileft)
       +(4,-2) coordinate (Iright)
       +(2,0) coordinate (bs)
       ;

\draw (Ibot) -- 
		node[xshift=1.5cm, yshift=-0.2cm, sloped] {\tiny $r=2M$, $t=\infty$} 
		node[xshift=1.5cm, yshift=0.2cm, sloped] {\scriptsize $\mathcal{H}^{\rm f}$} 
	(Ileft);
\draw (Itop) -- 
		node[xshift=1.5cm, yshift=0.2cm, sloped] {\tiny $r=2M$, $t=-\infty$}  
		node[xshift=1.5cm, yshift=-0.2cm, sloped] {\scriptsize $\mathcal{H}^{\rm p}$}
		(Iright);
\draw[->] (-2,0) -- node[left]{$T$} (-2, 1);
\draw[->] (-2,0) -- node[below]{$X$} (-1, 0);
\draw[line width=1pt] (0.1,2) to[out=-30, in=210] (3.9,2);
\draw[line width=1pt] (0.1,-2) to[out=30, in=-210] (3.9,-2);
\draw[latex-] (3.8,1.8) to[out=-30,in=180] (5,2.2) node[right]{$U=0$};
\draw[latex-] (3.8,-1.8) to[out=-30,in=200] (5,-1.5) node[right]{$V=0$};
\draw (4.2,1.6) node[right] {\tiny $r={\rm const.}$} to[out=215, in=145] (4.2,-1.6);
\draw[latex-] (0.5,1.8) to[out=45,in=180] (0.8,2) node[right]{\tiny $T=+\sqrt{1+X^2}$};
\draw[latex-] (0.5,-1.8) to[out=-45,in=180] (0.8,-2) node[right]{\tiny $T=-\sqrt{1+X^2}$};

\end{tikzpicture}
\caption{Kruskal extension of the Schwarzschild exterior $\mathcal{D}={\rm I}\backslash(\mathcal{H}^{\rm f}\cup\mathcal{H}^{\rm p})$. Each point in the diagram corresponds to a 2-sphere. The coordinates $(T,X)$ are $T:=\frac{V-U}{2}$, $X:=\frac{V+U}{2}$. The thick lines represent the singularity $T^2-X^2=1$.}
\label{Fig:Kruskal}
\end{figure}
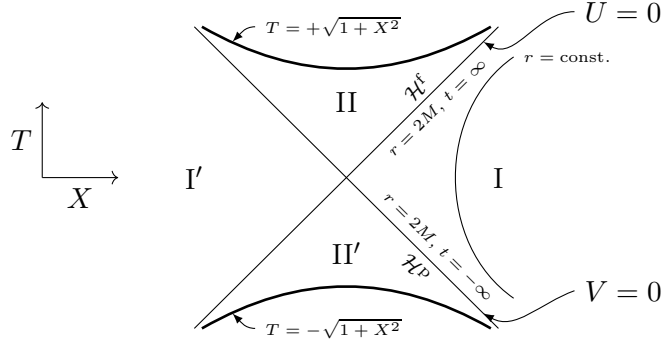

The metric \eqref{Kruskal} is regular at $r=2M$, so the spacetime can be extended across it. The condition \eqref{UVSchw} implies $U>0, V>0$, so we introduce an extended spacetime by defining the metric to be \eqref{Kruskal} but where now the coordinate values $U<0$ and $V<0$ are allowed:
\begin{align*}
 {\rm I}=\{U\geq0, V\geq0\}, \quad  {\rm II}=\{U\leq0, V\geq0\},  \quad {\rm I'}=\{U\leq0, V\leq0\}, \quad  {\rm II'}=\{U\geq0, V\leq0\}.
\end{align*}
The function $r(U,V)$ in the extended spacetime is still defined by \eqref{S-coords}, but now the possibility $r\leq2M$ is allowed. We see from \eqref{Kruskal} that the only problem is at $r=0$, i.e.,~at $UV=-1$. This is a curvature singularity and the spacetime cannot be extended beyond it; see fig. \ref{Fig:Kruskal}.

To describe the new regions ${\rm II, I', II'}$ from the point of view of Schwarzschild coordinates \eqref{Schw}, we can proceed analogously to the Rindler case.  
We define
\begin{align}
 U:= \begin{cases} e^{-\kappa u}, \quad {\rm I, II'}  \\ -e^{-\kappa u}, \quad {\rm I',II} \end{cases}, 
 \qquad
 V:= \begin{cases} e^{\kappa v}, \quad {\rm I,II}  \\ -e^{\kappa v}, \quad {\rm I',II'} \end{cases}
\end{align}
where $(u,v)$ are related to $(t,r)$ by \eqref{zSchw}.

{Although this may seem quite familiar so far, w}{e now wish to present a different perspective on the relations among the Schwarzschild times ``$t$'' of each of the four regions, that leads to  imaginary periodicity as in the Rindler case.}
We again introduce new (generally complex) null coordinates:
\bea
 \tilde{u} = t - r - 2M\,\Log\left( \frac{r}{2M} - 1 \right), \qquad 
 \tilde{v} = t+r + 2M\,\Log\left( \frac{r}{2M} - 1 \right).\label{tSchw}
\eea
Using $\rho=\frac{r}{2M}-1$, we again let $L=\Log({\rho})$ receive an imaginary $\delta L=i\pi$ jump when the sign of $\rho$ changes from positive to negative, on moving from region ${\rm I}$ into regions ${\rm II}$ or ${\rm II}'$.  As in the Rindler case, for $e^{\kappa\tilde{v}}$ to remain continuous across the future horizon $\mathcal{H}^{\rm f}$, and for $e^{-\kappa\tilde{u}}$ to remain continuous across the past horizon $\mathcal{H}^{\rm p}$ in passing out of region ${\rm I}$, it is then necessary that $t$ receive a $\delta t_{\rm I,II}=-\i\pi/(2\kappa)$ jump when passing into region ${\rm II}$, and a $\delta t_{\rm I,II'}=+\i\pi/(2\kappa)$ jump when passing into region ${\rm II'}$.  As a consequence, in region ${\rm II}$, $e^{-\kappa \tilde{u}}$ becomes $-e^{-\kappa u}={U|_{\rm II}}$, and in region ${\rm II'}$, $e^{\kappa \tilde{v}}$ becomes $-e^{\kappa v}={V|_{\rm II'}}$.  

Similarly, in passing from regions ${\rm II}$ or ${\rm II'}$ into region ${\rm I'}$, $L=\Log(\rho)$ will receive a relative $\delta L=-\i\pi$ jump, while $t$ will receive an additional $\delta t_{\rm II,I'}=-\i\pi/(2\kappa)$ jump when passing from region ${\rm II}$ into region ${\rm I'}$, so that $e^{\kappa \tilde{v}}$ becomes $-e^{\kappa v}={V|_{\rm I'}}$ while $e^{-\kappa \tilde{u}}$ remains $-e^{-\kappa u}={U|_{\rm I'}}$; alternatively, $t$ will receive an additional $\delta t_{\rm II',I'}=\i\pi/(2\kappa)$ jump when passing from region ${\rm II'}$ into region ${\rm I'}$, so that $e^{-\kappa \tilde{u}}$ becomes $-e^{-\kappa u}={U|_{\rm I'}}$ while $e^{\kappa \tilde{v}}$ remains $-e^{\kappa v}={V|_{\rm I'}}$.  

These last two results must be equivalent, so $t$ must have imaginary periodicity $\beta=2\pi/\kappa$ on the extended, Kruskal, manifold, i.e.:
\bea
t \sim t + \i\beta, \quad{\rm with}\quad \beta\equiv\frac{2\pi}{\kappa}.\label{S-per}
\eea 
Furthermore, with the understanding we have again developed for jumps in $t$ between neighbouring regions, ($\tilde{u},\tilde{v}$) become global, piecewise continuous coordinates, and the identifications:
\begin{align}
 \tilde{U} = e^{-\kappa \tilde{u}}, \qquad \tilde{V} = e^{\kappa \tilde{v}}, \label{UVS-t}
\end{align}
can be applied in all four regions simultaneously.

\section{Kerr}
\label{sec:Kerr}

\subsection{Preliminaries}

Consider, now, sub-extreme Kerr in Boyer-Lindquist coordinates \cite[eq. (2.13)]{BL} \footnote{Our ``$t$ '' and ``$\phi$ '' in \eqref{KerrBL} correspond to ``$\bar{t}$ '' and ``$\bar\varphi$ '' in \cite{BL}.}:
\begin{align}
 \d{s}^2 = g_{tt}\d{t}^2+2g_{t\phi}\d{t}\d\phi+g_{\phi\phi}\d\phi^2 + \frac{\Sigma}{\Delta}\d{r}^2 + 
 \Sigma\d\theta^2, 
 \label{KerrBL}
\end{align}
where 
\begin{gather}
  g_{tt} = \frac{a^2\sin^2\theta-\Delta}{\Sigma}, \quad
 g_{t\phi} = -\frac{2aMr\sin^2\theta}{\Sigma}, \quad
 g_{\phi\phi} =  \frac{[(r^2+a^2)^2 - \Delta a^2\sin^2\theta] \sin^2\theta}{\Sigma}, \\
   \Delta := r^2-2Mr+a^2, \qquad \Sigma := r^2+a^2\cos^2\theta,
\end{gather}
$M$ and $a$ are two real constants satisfying $M>|a|>0$, and
\begin{align}
  r_{\pm} = M \pm \sqrt{M^2-a^2} \label{rpm}
\end{align}
are the roots of $\Delta$. There are apparent singularities at $r=r_{\pm}$, but the spacetime can be extended across them. This was done by Boyer and Lindquist \cite{BL} and by Carter \cite{Carter1968}. Again the idea is to use coordinates based on null geodesics. 

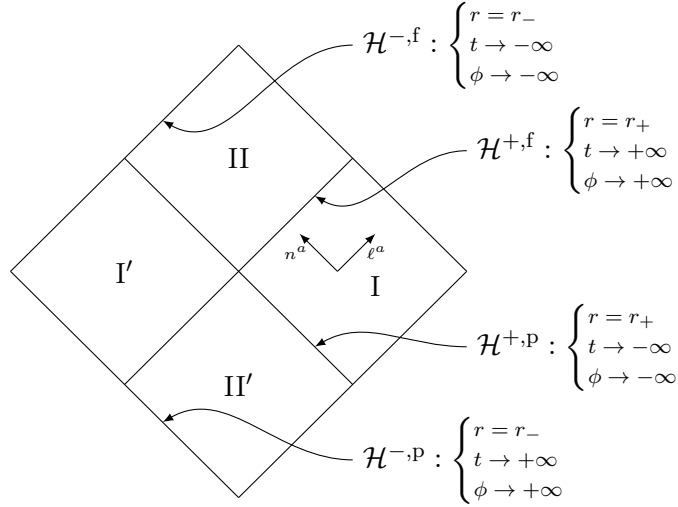
\begin{figure}
\centering
\begin{tikzpicture}
\draw (-3,0) -- (0,-3);
\draw (0,-3) -- (3,0);
\draw (3,0) -- (0,3);
\draw (0,3) -- (-3,0);
\draw (-1.5,-1.5) -- (1.5,1.5);
\draw (1.5,-1.5) -- (-1.5,1.5);
\node (I)    at (1.8,-0.2)   {I};
\node (I')   at (-1.5,0)   {I$'$};
\node (II)  at (0,1.5) {II};
\node (II')   at (0,-1.5) {II$'$};

\draw[latex-] (1,1) to[out=-30,in=180] (3,1.6) 
	node[right]{$\mathcal{H}^{+,{\rm f}}: {\scriptsize \begin{cases} r=r_{+} \\ t\to+\infty \\ 	
	\phi\to+\infty \end{cases} }$}; 
\draw[latex-] (1,-1) to[out=30,in=180] (3,-1) 
	node[right]{$\mathcal{H}^{+,{\rm p}}: {\scriptsize \begin{cases} r=r_{+} \\ t\to-\infty \\ 
	\phi\to-\infty \end{cases} }$};	
\draw[latex-] (-1,2) to[out=-30,in=180] (1.5,3) 
	node[right]{$\mathcal{H}^{-,{\rm f}}: {\scriptsize \begin{cases} r=r_{-} \\ t\to-\infty \\ 
	\phi\to-\infty \end{cases} }$}; 
\draw[latex-] (-1,-2) to[out=30,in=180] (1.5,-2.5) 
	node[right]{$\mathcal{H}^{-,{\rm p}}: {\scriptsize \begin{cases} r=r_{-} \\ t\to+\infty \\ 
	\phi\to+\infty \end{cases} }$}; 

\draw[-latex] (1.3,0) -- node[right]{\tiny $\ell^{a}$} (1.8,0.5);
\draw[-latex] (1.3,0) -- node[left]{\tiny $n^{a}$} (0.8,0.5);

\end{tikzpicture}

\caption{Behaviour of the Boyer-Lindquist coordinates $t,\phi$ near the future (${\rm f}$) and past (${\rm p}$) outer and inner horizons, as we approach them via the principal null congruences.}
\label{Fig:Kerr}
\end{figure}

As in Schwarzschild, there are preferred null geodesics adapted to the geometry: the PNDs $\ell^a$ and $n^a$. A suitable null tetrad aligned to the PNDs is \cite{Znajek}
\begin{subequations}\label{CarterTetrad}
\begin{align}
 \ell^{a}\partial_{a} &= \frac{1}{\sqrt{2\Sigma\Delta}}\left[(r^2+a^2)\partial_{t} + \Delta\partial_{r} + a\partial_{\phi}  \right], \label{lKerr} \\
  n^{a}\partial_{a} &= \frac{1}{\sqrt{2\Sigma\Delta}}\left[(r^2+a^2)\partial_{t} - \Delta\partial_{r} + a\partial_{\phi} \right], \label{nKerr} \\
  m^{a}\partial_{a} &= \frac{1}{\sqrt{2\Sigma}}\left[ \i a \sin\theta\partial_{t} + \partial_{\theta} + \frac{\i}{\sin\theta}\partial_{\phi} \right], \\
  \bar{m}^{a}\partial_{a} &= \frac{1}{\sqrt{2\Sigma}}\left[ -\i a \sin\theta\partial_{t} + \partial_{\theta} - \frac{\i}{\sin\theta}\partial_{\phi} \right].
\end{align}
\end{subequations}
{Integral curves of $\ell=\ell^a\partial_a$ are given by $\ell^a=(\dot{t},\dot{r},\dot\theta,\dot\phi)$, and we deduce from \eqref{lKerr}}
that $\dot{t}=\frac{(r^2+a^2)}{\sqrt{2\Sigma\Delta}}$, $\dot{r}=(\frac{\Delta}{2\Sigma})^{1/2}$, $\dot\theta=0$, $\dot\phi = \frac{a}{\sqrt{2\Sigma\Delta}}$. Analogous formula{e} follow from \eqref{nKerr} for $n^{a}$, the only difference is a $-$ sign in $\dot{r}$. Thus we deduce that the principal null congruences satisfy
\begin{align}
\frac{\d{t}}{\d{r}} = \pm \frac{(r^2+a^2)}{\Delta}, \qquad \frac{\d\phi}{\d{r}} = \pm \frac{a}{\Delta}, \qquad \theta = {\rm const.}
\end{align}
where the upper sign corresponds to $\ell^a$ and gives outgoing geodesics, and the lower sign corresponds to $n^{a}$ and gives ingoing geodesics. Integrating, we get
\begin{align}\label{tphiPNDs}
 \text{outgoing:} \,\, 
 \begin{cases}
 t = r^{*} + u \\
 \phi = r^{\sharp} + w \\
 \theta = {\rm const.}
 \end{cases},
 \qquad
 \text{ingoing:} \,\, 
 \begin{cases}
 t = -r^{*} + v \\
 \phi = -r^{\sharp} + z \\
 \theta = {\rm const.}
 \end{cases}
\end{align}
where $u,v$ and $w,z$ are constants, and we introduced
\begin{align}
 r^{*}(r):={}& \int\frac{(r^2+a^2)}{\Delta}\d{r} 
 = r + \frac{1}{2\kappa_{+}}\log\left| \frac{r}{r_{+}} -1 \right| 
 - \frac{1}{2\kappa_{-}}\log\left| \frac{r}{r_{-}} - 1 \right|, \\
 r^{\sharp}(r) :={}& a\int\frac{\d{r}}{\Delta} 
 = \frac{\Omega_{+}}{2\kappa_{+}}\log\left| \frac{r}{r_{+}} -1 \right| 
 - \frac{\Omega_{-}}{2\kappa_{-}}\log\left| \frac{r}{r_{-}} - 1 \right|,
\end{align}
where 
\begin{align}
 \kappa_{\pm} := \frac{r_{+}-r_{-}}{2(r_{\pm}^2+a^2)}, \qquad
  \Omega_{\pm} := \frac{a}{r_{\pm}^2+a^2}.
\end{align}
(We note that $\frac{\kappa_{+}}{\Omega_{+}}=\frac{\kappa_{-}}{\Omega_{-}}$.)
From \eqref{tphiPNDs} we deduce the behaviour of $t,\phi$ as we approach the horizons via the principal {null }congruences: near the outer horizon,
\begin{align}\label{tphir+}
 \text{outgoing:} \,\, 
 \begin{cases}
 t \to -\infty \quad \text{as} \quad r\to r_{+} \\
 \phi \to -\infty \quad \text{as} \quad  r\to r_{+}
 \end{cases},
 \qquad
 \text{ingoing:} \,\, 
 \begin{cases}
 t \to +\infty \quad \text{as} \quad r\to r_{+} \\
 \phi \to +\infty \quad \text{as} \quad r\to r_{+}
 \end{cases},
\end{align}
whereas near the inner horizon:
\begin{align}\label{tphir-}
 \text{outgoing:} \,\, 
 \begin{cases}
 t \to +\infty \quad \text{as} \quad r\to r_{-} \\
 \phi \to +\infty \quad \text{as} \quad  r\to r_{-}
 \end{cases},
 \qquad
 \text{ingoing:} \,\, 
 \begin{cases}
 t \to -\infty \quad \text{as} \quad r\to r_{-}, \\
 \phi \to -\infty \quad \text{as} \quad r\to r_{-}
 \end{cases}.
\end{align}
This is illustrated in the Penrose diagram of figure \ref{Fig:Kerr}.

\subsection{Extended coordinates {and {$\alpha-$}surfaces}}

In Schwarzschild, the procedure to cross the past and future horizons was to follow outgoing and ingoing geodesics, so we exchanged $(t,r)$ by $(u,v)$, and since the angular variables were untouched, we could work with the coordinate system $(u,v,\theta,\phi)$. In Kerr, we see from \eqref{tphiPNDs} that
\begin{align}\label{uwvz}
 \text{outgoing:} \,\, 
 \begin{cases}
 u = t - r^{*} \\
 w = \phi - r^{\sharp} \\
 \theta = {\rm const.}
 \end{cases},
 \qquad
 \text{ingoing:} \,\, 
 \begin{cases}
 v = t + r^{*} \\
 z = \phi + r^{\sharp} \\
 \theta = {\rm const.}
 \end{cases},
\end{align}
so we have too many variables: we can work with $(u,v,w,\theta)$, or with $(u,v,z,\theta)$, or similar combinations, but none of these will be simultaneously adapted to both the outgoing and ingoing congruences. Therefore we need to do something different.

We {thus }define the following complex scalar fields (see \cite{AA}, and Appendix \ref{Appendix} for details):
\begin{subequations}\label{LCCKerr}
\begin{align}
\hat{u} :={}& u -\i a \cos\theta, \\
 \hat{v} :={}& v + \i a \cos\theta, \\
\hat{w}:={}& w + \i \log\tan(\theta/2), \\
 \hat{z}:={}& z - \i \log\tan(\theta/2).
\end{align}
\end{subequations}
Since the  $\{\hat{u},\hat{v},\hat{w},\hat{z}\}$ are functionally independent, they define a complex coordinate system for Kerr. Moreover, the Kerr line element \eqref{KerrBL} becomes: 
\begin{align}
 \d{s}^2 = g_{tt}\d{\hat{u}}\d{\hat{v}}+ g_{t\phi}(\d{\hat{u}}\d{\hat{z}}+\d{\hat{v}}\d{\hat{w}}) + g_{\phi\phi}\d{\hat{w}}\d{\hat{z}}
\end{align} 
This system is especially adapted to the principal null congruences: comparing \eqref{LCCKerr} with \eqref{uwvz}, we see that if we fix an outgoing geodesic then we fix $(\hat{u},\hat{w})$ (assuming that $(t,r,\theta,\phi)$ are here real), while if we fix an ingoing geodesic then we fix $(\hat{v},\hat{z})$.

In the complexified Kerr spacetime, the set $(\hat{u},\hat{w})={\rm const.}$ describes an {\em $\alpha-$surface} {--} a {2D} complex surface such that all tangent vectors are null and orthogonal \cite{Penrose1976, PR2}. Such surfaces are also totally geodesic. Their existence is equivalent to the existence of a shear-free null geodesic congruence. There is a 2-complex-parameter system of $\alpha-$surfaces associated to the outgoing congruence, see \cite[Prop. (7.3.18)]{PR2}. Analogously, there is a second set of $\alpha-$surfaces associated to the ingoing congruence, where each of these surfaces is described by $(\hat{v},\hat{z})={\rm const}$.

The coordinates $\hat{u}$ and $\hat{w}$ are naturally paired by the structure of $\alpha-$surfaces, and analogously for $\hat{v}$ and $\hat{z}${. Moreo}{ver, one can perform holomorphic transformations of the form 
\begin{align}
(\hat{u},\hat{w}) \to (\hat{u}'(\hat{u},\hat{w}), \hat{w}'(\hat{u},\hat{w})), \qquad 
(\hat{v},\hat{z}) \to (\hat{v}'(\hat{v},\hat{z}), \hat{z}'(\hat{v},\hat{z})), \label{CCfreedom}
\end{align}
and the new coordinates $(\hat{u}',\hat{w}')$ and $(\hat{v}',\hat{z}')$ will describe the same families of $\alpha-$surfaces. That is, \eqref{CCfreedom} preserves the complex structure underlying our construction (see eq. \eqref{biholomorphism}).}

Let us now focus on the region ${\rm I}\cup{\rm II}\cup{\rm I'}\cup{\rm II'}$ (i.e., $r>r_-$). Using \eqref{LCCKerr}, which are separately defined in all four regions, we define
\begin{subequations}\label{GCCKerr}
\begin{align}
 U_{+} &:= \begin{cases} 
 	\exp(-\kappa_{+}\hat{u}), \qquad & {\rm I, II'} \\ 
	- \exp(-\kappa_{+}\hat{u}), \qquad & {\rm I', II} 
	\end{cases} 
\qquad & V_{+} &:= \begin{cases}
 	 \exp(\kappa_{+}\hat{v}), \qquad & {\rm I, II} \\ 
	 - \exp(\kappa_{+}\hat{v}), \qquad & {\rm I', II'} 
	 \end{cases} \\
 W_{+} &:= \begin{cases} 
 	\exp(-\tfrac{\kappa_{+}}{\Omega_{+}}\hat{w}), \qquad & {\rm I, II'} \\ 
	- \exp(-\tfrac{\kappa_{+}}{\Omega_{+}}\hat{w}), \qquad & {\rm I', II} 
	\end{cases} 
\qquad & Z_{+} &:= \begin{cases} 
 	\exp(\tfrac{\kappa_{+}}{\Omega_{+}}\hat{z}), \qquad & {\rm I, II} \\ 
	- \exp(\tfrac{\kappa_{+}}{\Omega_{+}}\hat{z}), \qquad & {\rm I', II'} 
	\end{cases}
\end{align}
\end{subequations}
To gain some intuition, we can write these variables in region ${\rm I}$:
\begin{subequations}
\begin{align}
 U_{+}\big|_{\rm I} &= e^{-\kappa_{+}(t-r-\i a \cos\theta)}\left(\frac{r}{r_{+}}-1\right)^{1/2}\left(\frac{r}{r_{-}}-1\right)^{-\frac{\kappa_{+}}{2\kappa_{-}}}, \\
  V_{+}\big|_{\rm I} &= e^{\kappa_{+}(t+r +\i a \cos\theta)}\left(\frac{r}{r_{+}}-1\right)^{1/2}\left(\frac{r}{r_{-}}-1\right)^{-\frac{\kappa_{+}}{2\kappa_{-}}},  \\
 W_{+}\big|_{\rm I} &= e^{-\frac{\kappa_{+}}{\Omega_{+}}(\phi+\i\log\tan\frac{\theta}{2})}\left(\frac{r}{r_{+}}-1\right)^{1/2}\left(\frac{r}{r_{-}}-1\right)^{-1/2}, \\
 Z_{+}\big|_{\rm I} &= e^{\frac{\kappa_{+}}{\Omega_{+}}(\phi - \i\log\tan\frac{\theta}{2})}\left(\frac{r}{r_{+}}-1\right)^{1/2}\left(\frac{r}{r_{-}}-1\right)^{-1/2}.
\end{align}
\end{subequations}
From the behaviour \eqref{tphir+} (see also fig. \ref{Fig:Kerr}), we deduce
\begin{subequations}\label{horizons}
\begin{align}
  U_{+}\big|_{\mathcal{H}^{+, {\rm f}}} &= 0 , \label{Ufh} \\ 
  W_{+}\big|_{\mathcal{H}^{+, {\rm f}}} &= 0, \label{Wfh} \\
  V_{+}\big|_{\mathcal{H}^{+, {\rm p}}} &= 0, \label{Vph} \\ 
  Z_{+}\big|_{\mathcal{H}^{+, {\rm p}}} &= 0, \label{Zph}
\end{align}
\end{subequations}
and each is continuous across its respective horizon.  Hence, $(U_{+},V_{+},W_{+},Z_{+})$ cover ${\rm I}\cup{\rm II}\cup{\rm I'}\cup{\rm II'}$. 

\smallskip

In terms of the variables \eqref{GCCKerr}, one family of $\alpha-$surfaces is now described by $(U_+,W_+) = const.$, and the other family of $\alpha-$surfaces is described by $(V_+,Z_+) = const.$ (recall \eqref{CCfreedom}). Since we see from \eqref{Ufh} and \eqref{Wfh} that at $\mathcal{H}^{+, {\rm f}}$ we have $(U_+,W_+)=(0,0)$, then the future horizon lies in an $\alpha-$surface. The points of $\mathcal{H}^{+, {\rm f}}$ are coordinatized by the two complex coordinates $(V_+,Z_+)$. One might think that there is an inconsistency here since $\mathcal{H}^{+, {\rm f}}$ is a {\em three}-real-dimensional hypersurface and two complex coordinates generically amount to {\em four} real coordinates, however there is no  
problem {here }since we see from \eqref{GCCKerr} and \eqref{LCCKerr} that the imaginary parts of $V_+$ and $Z_+$ are functionally dependent (they are both functions of $\theta$). In real terms, $\mathcal{H}^{+, {\rm f}}$ is coordinatized by $(v,z,\theta)$. Analogously, we see from \eqref{Vph}-\eqref{Zph} that the past horizon $\mathcal{H}^{+, {\rm p}}$ lies in an $\alpha-$surface (of the other family), and it is coordinatized by the two complex coordinates $(U_+,W_+)$ or by the three real coordinates $(u,w,\theta)$.

{I}t is not clear from \eqref{horizons} what happens at the intersection of the future and past horizons, i.e. at the bifurcation sphere. We will address this point in the next subsection, after we {have discussed} imaginary shifts{, now in both $t$ and $\phi$,} across {the} horizons.

\subsection{Imaginary shifts, periodicity, and the bifurcation sphere}

To proceed as in the previous examples, we now introduce coordinates obtained by using $\Log$ rather than $\log$ in the definitions of new radial functions $\widetilde{r^*}$ and $\widetilde{r^{\sharp}}$:
\begin{align}
 \widetilde{r^{*}}(r):={}& r + \frac{1}{2\kappa_{+}}\Log\left( \frac{r}{r_{+}} -1 \right) 
 - \frac{1}{2\kappa_{-}}\Log\left( \frac{r}{r_{-}} - 1 \right), \\
 \widetilde{r^{\sharp}}(r) :={}& \frac{\Omega_{+}}{2\kappa_{+}}\Log\left( \frac{r}{r_{+}} -1 \right) 
 - \frac{\Omega_{-}}{2\kappa_{-}}\Log\left( \frac{r}{r_{-}} - 1 \right).
\end{align}
Then, our new tilde-coordinates will become:
\begin{align}
 \tilde{u} :={}& t - \widetilde{r^{*}} -\i a \cos\theta, \\
  \tilde{v} :={}& t+\widetilde{r^{*}} + \i a \cos\theta, \\
 \tilde{w}:={}& \phi - \widetilde{r^{\sharp}} + \i \log\tan(\theta/2), \\
  \tilde{z}:={}& \phi + \widetilde{r^{\sharp}} - \i \log\tan(\theta/2).
\end{align}
As in the previous examples, $L=\Log(r/r_+-1)$ will receive a $\delta L=i\pi$ jump on moving from region ${\rm I}$ into region ${\rm II}$ or ${\rm II}'$.  However, now we will have to consider all four coordinates, instead of just two.  So, for $e^{\kappa_+\tilde{v}}$, and now also $e^{\kappa_+\tilde{z}/\Omega_+}$, to remain continuous across the future horizon ${\mathcal{H}^{+,{\rm f}}}$, and for $e^{-\kappa_+\tilde{u}}$, and now also $e^{\kappa_+\tilde{w}/\Omega_+}$, to remain continuous across the past horizon ${\mathcal{H}^{+,{\rm p}}}$ in passing out of region ${\rm I}$, it is again necessary that $t$ receive a $\delta t_{\rm I,II}=-\i\pi/(2\kappa_+)$ jump when passing into region ${\rm II}$, and a $\delta t_{\rm I,II'}=+\i\pi/(2\kappa_+)$ jump when passing into region ${\rm II'}$, while $\phi$ must receive a jump $\delta\phi=\Omega_+\delta t$ in both cases.  As a consequence, in region ${\rm II}$, $e^{-\kappa_+ \tilde{u}}$ becomes $-e^{-\kappa_+ \hat{u}}=U_{+}|_{\rm II}$ and $e^{-\kappa_+ \tilde{w}/\Omega_+}$ becomes $-e^{-\kappa_+ \hat{w}/\Omega_+}=W_{+}|_{\rm II}$, while in region ${\rm II'}$, $e^{\kappa_+ \tilde{v}}$ becomes $-e^{\kappa_+ \hat{v}}=V_{+}|_{\rm II'}$ and $e^{\kappa_+ \tilde{z}/\Omega_+}$ becomes $-e^{\kappa_+ \hat{z}/\Omega_+}=Z_{+}|_{\rm II'}$.  

Similarly, in passing from regions ${\rm II}$ or ${\rm II'}$ into region ${\rm I'}$, $L=\Log(r/r_+-1)$ will receive a relative $\delta L=-\i\pi$ jump, while $t$ will receive an additional $\delta t_{\rm II,I'}=-\i\pi/(2\kappa_+)$ jump, and $\phi$ will receive an additional $\delta\phi=\Omega_+\delta t$ jump, when passing from region ${\rm II}$ into region ${\rm I'}$, so that $e^{\kappa_+ \tilde{v}}$ becomes $-e^{\kappa_+ \hat{v}}=V_{+}|_{\rm I'}$ and $e^{\kappa_+ \tilde{z}/\Omega_+}$ becomes $-e^{\kappa_+ \hat{z}/\Omega_+}=Z_{+}|_{\rm I'}$, while $e^{-\kappa_+ \tilde{u}}$ remains $-e^{-\kappa_+ \hat{u}}=U_{+}|_{\rm I'}$ and $e^{-\kappa_+ \tilde{w}/\Omega_+}$ remains $-e^{-\kappa_+ \hat{w}/\Omega_+}=W_{+}|_{\rm I'}$; alternatively, $t$ will receive an additional $\delta t_{\rm II',I'}=\i\pi/(2\kappa_+)$ jump, and $\phi$ will receive an additional $\delta\phi=\Omega_+\delta t$ jump,  when passing from region ${\rm II'}$ into region ${\rm I'}$, so that $e^{-\kappa_+ \tilde{u}}$ becomes $-e^{-\kappa_+ \hat{u}}=U_{+}|_{\rm I'}$ and $e^{-\kappa_+ \tilde{w}/\Omega_+}$ becomes $-e^{-\kappa_+ \hat{w}/\Omega_+}=W_{+}|_{\rm I'}$, while $e^{\kappa_+ \tilde{v}}$ remains $-e^{\kappa_+ \hat{v}}=V_{+}|_{\rm I'}$ and $e^{\kappa_+ \tilde{z}/\Omega_+}$ remains $-e^{\kappa_+ \hat{z}/\Omega_+}=Z_{+}|_{\rm I'}$.  These last two sets of results concerning $t$ and $\phi$ in region ${\rm I}'$ must be equivalent, that is, points obtained going anti-clockwise around the diagram and points obtained by going clockwise around the diagram must be identified, i.e., for $\beta_+=2\pi/\kappa_+$:
\bea
\left(t-\frac{i\beta_+}{2},\phi-\frac{i\beta_+}{2}\Omega_+\right)&\sim&\left(t+\frac{i\beta_+}{2},\phi+\frac{i\beta_+}{2}\Omega_+\right).
\label{identifK1}
\eea
We can phrase this outcome more explicitly in terms of periodicity by defining the new quantity $\Phi_+=\phi-\Omega_+ t$.  Then:
\bea
(t,\Phi_+)\sim(t,\Phi_+ +2\pi)\sim(t+i\beta_+,\Phi_+)\sim(t+i\beta_+,\Phi_+ +2\pi),
\label{identifK2}
\eea
so $t$ must have imaginary periodicity $\beta_+=2\pi/\kappa_+$ on the extended Kerr manifold, and $\phi$ must also have a total imaginary shift of $\Omega_+\beta_+$.
Furthermore, with the understanding we have again developed for jumps in $t$, 
{and now also in $\phi$,} between neighbouring regions, ($\tilde{u},\tilde{v},{\tilde{w},\tilde{z}}$) become global, piecewise continuous coordinates, and the identifications:
\begin{align}
 \tilde{U}_+ = e^{-\kappa_+ \tilde{u}}, \qquad \tilde{V}_+ = e^{\kappa_+ \tilde{v}}, 
 \qquad 
 {\tilde{W}_+ = e^{-\kappa_+ \tilde{w}/\Omega_+}, \qquad  \tilde{Z}_+ = e^{\kappa_+ \tilde{z}/\Omega_+}}
 \label{UVK-t}
\end{align}
can be applied in all four regions simultaneously.

To address the issue of the bifurcation sphere, we introduce the following two coordinates:
\bea
{\check w_{+}}&=&\tilde{w}-\Omega_+\tilde{u},\\
{\check z_{+}}&=&\tilde{z}-\Omega_+\tilde{v}.
\eea
Note that ${\check w_{+}}$ and ${\check z_{+}}$ respect the structure of $\alpha-$surfaces (recall \eqref{CCfreedom}). We have:
\begin{align}
{\check w_{+}} ={}& \Phi_{+} + \frac{(\Omega_{-}-\Omega_{+})}{2\kappa_{-}}\Log\left(\frac{r}{r_{-}}-1\right) + \i \, (\log\tan(\theta/2)+a\Omega_{+}\cos\theta), \\
{\check z_{+}} ={}& \Phi_{+} - \frac{(\Omega_{-}-\Omega_{+})}{2\kappa_{-}}\Log\left(\frac{r}{r_{-}}-1\right) - \i \, (\log\tan(\theta/2)+a\Omega_{+}\cos\theta).
\end{align}
Thus, we see that ${\check w_{+}}$ and ${\check z_{+}}$ are regular at the bifurcation sphere. We can then use the coordinate system $(U_{+},V_{+},{\check w_{+}},{\check z_{+}})$ to cover the entire region outside the inner horizon, and the bifurcation sphere corresponds to $U_{+}=0,V_{+}=0$ and is coordinatized by $({\check w_{+}},{\check z_{+}})$.

\smallskip
{We have focused on the outer horizon $r_{+}$, but an entirely analogous description can be developed for the inner horizon $r_{-}$, using the local complex coordinate system \eqref{LCCKerr} to define the analogs of \eqref{GCCKerr} but now replacing $\kappa_{+},\Omega_{+}$ by $\kappa_{-},\Omega_{-}$. This does not introduce any complications, so we will omit the details here.}

\section{Discussion}
\label{sec:Discussion}

The global structure of a maximally extended black hole spacetime is separated into a series of regions bounded by horizons, each region having a local ``time'' coordinate $t$ and ``azimuthal angle'' $\phi$ (more precisely, $t$ and $\phi$ parametrize the orbits of the Killing fields that are respectively timelike and spacelike ---the latter with closed orbits--- in the asymptotic infinity of the black hole exterior). In this paper, by exploiting the special null geodesic structure as understood from the perspective of $\alpha-$surfaces, we constructed coordinate systems that are continuous across horizons, cover the bifurcation sphere, and allow to show that the $t$'s and $\phi$'s in the different regions of rotating black holes are related by imaginary translations, while the horizons themselves are contained in $\alpha-$surfaces. This leads directly to imaginary time periodicity and similar shifts for $\phi$, without invoking Wick rotations, Euclidean sections, {n}or an artificial replacement of the angular momentum parameter.

\smallskip
{Note that, in a somewhat different perspective, we can also interpret the imaginary translations in local Schwarzschild or Kerr coordinates as the imposition of different reality conditions in the complexifications of those coordinates. In other words, this point of view leads to a curious picture in which the different regions of a global black hole spacetime are different real slices of its holomorphic extension. (Similar ideas about imaginary translations and reality conditions are presented when describing Kerr as a ---non-holomorphic--- complex translation of Schwarzschild, according to the Newman-Janis shift \cite{Flaherty, NewmanJanis}.) However, the above description is not exactly equivalent to the procedure we followed in this work, since it does not specify an unambiguous choice of analytic continuation across horizons, which is essential in our approach.}

\smallskip
Apart from the initial example of the {2D} Rindler spacetime, our discussion was centered around the extensions of the Schwarzschild and Kerr solutions. The horizon in the Rindler case is an acceleration horizon, whereas in the Schwarzschild and Kerr geometries it is an event horizon (including also a Cauchy horizon in the Kerr case). Since the key feature underlying our procedure is the existence of $\alpha-$surfaces, we can also apply our method to the other types of horizons that occur in spacetimes admitting $\alpha-$surfaces, such as the cosmological horizons of asymptotically de Sitter black holes, or the more complicated acceleration horizons of the C-metric and Pleba\'nski-Demia\'nski solutions.

\smallskip
In the original works assigning an entropy and a temperature to black holes \cite{Bekenstein:1973ur,Hawking:1975vcx}, there was no reference to Euclideanization.  
{However, in the subsequent development of Euclidean quantum gravity, thermodynamical aspects of the gravitational field were often studied by employing positive-definite sections (e.g. \cite{GH1979, GibbonsPerry, Lapedes:1980st, Caldarelli}), and this sometimes involved unnatural complexifications of the parameters of a solution.  By contrast, our use of $\alpha-$surfaces in this paper, which exposes a periodicity in imaginary $t$ and $\phi$, directly supports a thermodynamic interpretation \cite{GH1977,Witten};  that is, our method applies to complexified geometries, regardless of whether they admit real sections with a particular signature.  
In addition to the classical black holes of general relativity, our method of using $\alpha-$surfaces also applies to Lorentzian black holes that are not algebraically special, such as certain supergravity solutions (see \cite{AA}), and also gravitational instantons with Hermitian structures that do not have Lorentzian sections, but can still have physical effects (see \cite{Lapedes:1980st}), such as the Chen-Teo instanton \cite{ChenTeo1}, whose contribution to the quantum gravity partition function is as yet unknown.}

\smallskip
Finally, the special complex coordinate systems that we constructed for Kerr will allow us to study various aspects of classical and quantum fields on the entire region outside the inner horizon of rotating black holes, analogously to what has been done in the static case for the extended Schwarzschild \cite{SW} and Reissner-Nordstr\"om \cite{SWF} solutions. This includes the construction of analytic, globally positive or negative frequency fields, that will in turn be naturally related to a thermodynamical description. This will be presented in a separate work.

\paragraph{Acknowledgements.}
The authors acknowledge support of the Institut Henri Poincar\'e (UAR 839 CNRS-Sorbonne Universit\'e), and LabEx CARMIN (ANR-10-LABX-59-01).

\appendix

\section{Complex structures}
\label{Appendix}

In this appendix we summarize some properties about integrable complex structures in Lorentz signature, and we give further details for the Kerr case. We follow \cite{Flaherty} and \cite{AA}, with the exception that we use the mostly plus convention for the metric signature, $(-+++)$.

\paragraph{Lorentzian complex structures.} 
Let $(M,g_{ab})$ be a four-dimensional Lorentzian manifold. Consider a null tetrad $(\ell^a,n^a,m^a,\bar{m}^{a})$, with $g_{ab}\ell^{a}n^{b}=-1$, $g_{ab}m^{a}\bar{m}^{b}=1$. The dual tetrad is $(\ell_a,n_a,m_a,\bar{m}_{a})$. Define the tensor field 
\begin{align}
 J^{a}{}_{b} := \i(\ell^{a}n_{b} - n^{a}\ell_{b} + m^{a}\bar{m}_{b} - \bar{m}^{a}m_{b}). \label{J}
\end{align}
Then $J^{a}{}_{c}J^{c}{}_{b} = - \delta^{a}_{b}$ and $g_{ab}J^{a}{}_{c}J^{b}{}_{d}=g_{cd}$, so $J^{a}{}_{b}$ is an almost-complex structure compatible with $g_{ab}$. The tangent bundle decomposes as $TM\otimes\mathbb{C}=T^{+}\oplus T^{-}$, where $T^{\pm}$ are the eigenspaces of $J^{a}{}_{b}$ corresponding to the eigenvalue $\pm \i$. One can check that 
\begin{align}
 J^{a}{}_{b}\ell^{b} = - \i \, \ell^{a}, \qquad 
  J^{a}{}_{b}n^{b} = + \i \, n^{a},  \qquad J^{a}{}_{b}m^{b} = + \i \, m^{a},  \qquad J^{a}{}_{b}\bar{m}^{b} = - \i \, \bar{m}^{a},
\end{align}
so $T^{+}={\rm span}(n^a, m^{a})$, $T^{-}={\rm span}(\ell^a,\bar{m}^{a})$. We say that $J^{a}{}_{b}$ is integrable iff $[T^{+},T^{+}]\subset T^{+}$ and $[T^{-},T^{-}]\subset T^{-}$. If this is satisfied, there exist functions $A^{i},B^{i},\tilde{A}^{i},\tilde{B}^{i}$, with $i=0,1$, such that 
\begin{align}
A^{i}\ell_{a}\d{x}^{a} + B^{i}\bar{m}_{a}\d{x}^{a}, \qquad 
\tilde{A}^{i}n_{a}\d{x}^{a} + \tilde{B}^{i} m_{a}\d{x}^{a}
\end{align}
are closed 1-forms, and thus locally exact{.  T}here are {then }complex scalar fields $z^{i},\tilde{z}^{i}$ such that 
\begin{align}
 A^{i}\ell_{a}\d{x}^{a} + B^{i}\bar{m}_{a}\d{x}^{a} = \d{z}^{i}, \qquad 
\tilde{A}^{i}n_{a}\d{x}^{a} + \tilde{B}^{i} m_{a}\d{x}^{a} = \d\tilde{z}^{i}.
\label{ComplexCoordGeneral}
\end{align}
Denoting $g_{i\tilde{j}}:=g(\partial_{z^{i}},\partial_{\tilde{z}^{j}})$, with $i,j = 0,1$, the metric is
\begin{align}
 \d{s}^2 = 2 \, ( \, g_{0\tilde{0}} \d{z}^{0} \d\tilde{z}^{0} + g_{0\tilde{1}} \d{z}^{0} \d\tilde{z}^{1} + g_{1\tilde{0}} \d{z}^{1} \d\tilde{z}^{0} + g_{1\tilde{1}}\d{z}^{1}\d\tilde{z}^{1} \, ),
\end{align}
and the complex structure \eqref{J} becomes
\begin{align}
 J = \i(\partial_{z^0}\otimes\d{z}^0 + \partial_{z^1}\otimes\d{z}^1 
 - \partial_{\tilde{z}^0}\otimes\d\tilde{z}^0 - \partial_{\tilde{z}^1}\otimes\d\tilde{z}^1). \label{J2}
\end{align}
Note that \eqref{J2} is preserved by any transformation of the form
\begin{align}
 z^{i} \to z'^{i}=f^{i}(z^0,z^1), \qquad \tilde{z}^{i} \to \tilde{z}'^{i}=\tilde{f}^{i}(\tilde{z}^0,\tilde{z}^1), 
 \label{biholomorphism}
\end{align}
where $f^{i}$ and $\tilde{f}^{i}$ are any holomorphic functions of their arguments.

\paragraph{Shear-free congruences.} 
The integrability condition for $J^{a}{}_{b}$ turns out to be equivalent to the condition that $\ell^a$ and $n^{a}$ define shear-free null geodesic congruences. For a spacetime satisfying the vacuum Einstein equations, the Goldberg-Sachs theorem then implies that an integrable $J^{a}{}_{b}$ exists if and only if the spacetime is Petrov type D. In this case, the corresponding null tetrad that defines $J^{a}{}_{b}$ via \eqref{J} is given by the principal null directions (PNDs). To find the complex coordinates $z^{i},\tilde{z}^{i}$ in \eqref{ComplexCoordGeneral}, one must then use a null tetrad aligned to the PNDs.

\paragraph{The Kerr spacetime.} 
For Kerr, a null tetrad aligned to the PNDs was given in \eqref{CarterTetrad}. The dual tetrad is:
\begin{subequations}\label{DualCarterTetrad}
\begin{align}
\ell_{a}\d{x}^{a} &= \sqrt{\frac{\Delta}{2\Sigma}} \left[ - \d{t} +\frac{\Sigma}{\Delta} \, \d{r} 
+ a\sin^2\theta \, \d\phi \right], \label{ldKerr} \\
 n_{a}\d{x}^{a} &= \sqrt{\frac{\Delta}{2\Sigma}} \left[ - \d{t} - \frac{\Sigma}{\Delta} \, \d{r} 
 + a\sin^2\theta \, \d\phi \right], \label{ndKerr} \\
 m_{a}\d{x}^{a} &= \frac{1}{\sqrt{2\Sigma}} \left[ -\i a \sin\theta \, \d{t} + \Sigma \, \d\theta + \i\sin\theta \, (r^2+a^2) \, \d\phi \right], \label{mdKerr} \\
  \bar{m}_{a}\d{x}^{a} &= \frac{1}{\sqrt{2\Sigma}} \left[ + \i a \sin\theta \, \d{t} + \Sigma \, \d\theta - \i\sin\theta \, (r^2+a^2) \, \d\phi \right]. \label{mbdKerr} 
\end{align}
\end{subequations}
To find complex coordinates for Kerr, we use \eqref{ComplexCoordGeneral} with \eqref{DualCarterTetrad} and 
\begin{equation}
\begin{aligned}
 & A^{0} = -(r^2+a^2)\sqrt{\tfrac{2}{\Sigma\Delta}}, \qquad 
 B^{0} = +\i a\sin\theta\sqrt{\tfrac{2}{\Sigma}}, \qquad
 A^{1} = -a\sqrt{\tfrac{2}{\Sigma\Delta}}, \qquad 
 B^{1} = \tfrac{\i}{\sin\theta}\sqrt{\tfrac{2}{\Sigma}}, \\
 & \tilde{A}^{0} = A^{0}, \qquad 
 \tilde{B}^{0} = -B^{0}, \qquad 
 \tilde{A}^{1} = A^{1}, \qquad 
 \tilde{B}^{1} = -B^{1}.
\end{aligned}
\end{equation}
We also find $g_{0\tilde{0}} = g_{tt}\,$, $g_{0\tilde{1}} = g_{1\tilde{0}} = g_{t\phi}\,$, $g_{1\tilde{1}} = g_{\phi\phi}$. To compare the notation in this appendix to that of section \ref{sec:Kerr} (see eq. \ref{LCCKerr}), we {must }set
\begin{align}
 z^0 \equiv \hat{u}, \qquad \tilde{z}^0 \equiv \hat{v}, \qquad  z^1 \equiv \hat{w}, \qquad \tilde{z}^1 \equiv \hat{z}.
\end{align}


\begin{thebibliography}{99}

\bibitem{GHP1978}
G.~W.~Gibbons, S.~W.~Hawking and M.~J.~Perry,
{\em Path Integrals and the Indefiniteness of the Gravitational Action},
Nucl. Phys. B \textbf{138} (1978) 141-150

\bibitem{GHbook}
G.~W.~Gibbons and S.~W.~Hawking, 
{\em Euclidean quantum gravity}, (World Scientific, Singapore,1993) 600pp

\bibitem{Hawking1979}
S.~W.~Hawking, 
{\em The path-integral approach to quantum gravity}, 746-789 
in: General relativity: an Einstein centenary survey, 
(Cambridge University Press, Cambridge, 1979) 919pp

\bibitem{GH1977}
G.~W.~Gibbons and S.~W.~Hawking,
{\em Action Integrals and Partition Functions in Quantum Gravity},
Phys. Rev. D \textbf{15} (1977) 2752-2756

\bibitem{Hawking1976}
S.~W.~Hawking,
{\em Gravitational Instantons},
Phys. Lett. A \textbf{60} (1977) 81-83

\bibitem{GH1979}
G.~W.~Gibbons and S.~W.~Hawking,
{\em Classification of Gravitational Instanton Symmetries},
Commun. Math. Phys. \textbf{66} (1979) 291-310

\bibitem{Zee}
A.~Zee, 
{\em Quantum Field Theory in a Nutshell}, (2nd ed.), (Princeton University Press, Princeton, 2010) 608pp

\bibitem{Witten}
E.~Witten,  
{\em A note on complex spacetime metrics}, 
In Frank Wilczek: 50 years of theoretical physics (2022), pp. 245-280, 
\href{https://arxiv.org/abs/2111.06514}{[arXiv:2111.06514 [hep-th]]}

\bibitem{Brown}
J.~D.~Brown, E.~A.~Martinez and J.~W.~York, Jr.,
{\em Complex Kerr-Newman geometry and black hole thermodynamics},
Phys. Rev. Lett. \textbf{66} (1991) 2281-2284

\bibitem{Flaherty0}
 E.~J.~Flaherty Jr,  
 {\em An integrable structure for type D spacetimes}, 
 Physics Letters A, \textbf{46} (1974) 391-392

\bibitem{Flaherty}
 E.~J.~Flaherty Jr., 
 {\em Hermitian and K\"ahlerian Geometry in Relativity}, 
 Springer Lecture Notes in Physics, Vol. 46 (Springer-Verlag, 
 New York, 1976) 365pp

\bibitem{Penrose1976} 
  R.~Penrose,
  {\em Nonlinear Gravitons and Curved Twistor Theory},
  Gen.\ Rel.\ Grav.\  {\bf 7} (1976) 31-52

\bibitem{PR2}
  R.~Penrose and W.~Rindler,
  {\em Spinors And Space-time. Vol. 2: Spinor And Twistor Methods In Space-time Geometry},  (Cambridge University Press, Cambridge, 1986) 501pp

\bibitem{AA}
S.~Aksteiner and B.~Araneda,
{\em K\"ahler Geometry of Black Holes and Gravitational Instantons},
Phys. Rev. Lett. \textbf{130} (2023) no.16, 161502
\href{https://arxiv.org/abs/2207.10039}{[arXiv:2207.10039 [gr-qc]].}

\bibitem{PD}
 J.~F.~Pleba\'nski and M.~Demia\'nski,  
 {\em Rotating, charged, and uniformly accelerating mass in general 
 relativity}, Annals of Physics, \textbf{98} (1976) 98-127

\bibitem{Wald}
R.~M.~Wald, 
{\em General Relativity}, 
(Chicago University Press, Chicago, 1984) 491pp

\bibitem{BL}
R.~H.~Boyer and R.~W.~Lindquist,
{\em Maximal analytic extension of the Kerr metric},
J. Math. Phys. \textbf{8} (1967) 265-281

\bibitem{Carter1968}
B.~Carter,
{\em Global structure of the Kerr family of gravitational fields},
Phys. Rev. \textbf{174} (1968) 1559-1571

\bibitem{Znajek}
R.~Znajek,
{\em Black hole electrodynamics and the Carter tetrad},
Mon. Not. Roy. Astron. Soc. \textbf{179} (1977) 457-472

\bibitem{NewmanJanis}
E.~T.~Newman and A.~I.~Janis,
{\em Note on the Kerr spinning particle metric},
J. Math. Phys. \textbf{6} (1965) 915-917

\bibitem{Bekenstein:1973ur}
J.~D.~Bekenstein,
{\em Black holes and entropy},
Phys. Rev. D \textbf{7}, 2333-2346 (1973)

\bibitem{Hawking:1975vcx}
S.~W.~Hawking,
{\em Particle Creation by Black Holes},
Commun. Math. Phys. \textbf{43}, 199-220 (1975)
[erratum: Commun. Math. Phys. \textbf{46}, 206 (1976)]

\bibitem{GibbonsPerry}
G.~W.~Gibbons and M.~J.~Perry,
{\em Black Holes and Thermal Green's Functions},
Proc. Roy. Soc. Lond. A \textbf{358} (1978), 467-494


\bibitem{Lapedes:1980st}
A.~S.~Lapedes,
{\em Black Hole Uniqueness Theorems in Euclidean Quantum Gravity},
Phys. Rev. D \textbf{22} (1980) 1837-1847

\bibitem{Caldarelli}
M.~M.~Caldarelli, G.~Cognola and D.~Klemm,
{\em Thermodynamics of Kerr-Newman-AdS black holes and conformal field theories},
Class. Quant. Grav. \textbf{17} (2000), 399-420
\href{https://arxiv.org/abs/hep-th/9908022}{[arXiv:hep-th/9908022 [hep-th]].}

\bibitem{ChenTeo1}
Y.~Chen and E.~Teo,
{\em A New AF gravitational instanton},
Phys. Lett. B \textbf{703} (2011), 359-362
\href{https://arxiv.org/abs/1107.0763}{[arXiv:1107.0763 [gr-qc]]}.


\bibitem{SW}
N.~G.~Sanchez and B.~F.~Whiting,
{\em Quantum Field Theory and the Antipodal Identification of Black Holes},
Nucl. Phys. B \textbf{283} (1987) 605-623

\bibitem{SWF}
N.~A.~Strauss, B.~F.~Whiting and A.~T.~Franzen,
{\em Classical tools for antipodal identification in Reissner\textendash{}Nordstr\"om spacetime},
Class. Quant. Grav. \textbf{37} (2020) no.18, 185006
\href{https://arxiv.org/abs/2002.02501}{[arXiv:2002.02501 [gr-qc]].}

\end{thebibliography}
\end{document}